\documentstyle[12pt,a4wide]{article}
\input{psfig.tex}
\begin{document}
\renewcommand{\thefootnote}{\fnsymbol{footnote}}
\begin{flushright}
Imperial/TP/94-95/16\\
hep-ph/9502209\\
$31^{st}$ January 1995
\end{flushright}
\vskip 1cm
\begin{center}
{\Large\bf Thermalisation of Gauge Bosons in the Abelian Higgs Model}
\vskip 1.2cm
{\large\bf T.S.Evans\footnote{E-Mail: T.Evans@IC.AC.UK} \&
A.C.Pearson\footnote{E-mail: A.Pearson@IC.AC.UK}}\\
Blackett Laboratory, Imperial College, Prince Consort Road,\\
London SW7 2BZ  U.K. \\
\end{center}

\vskip 1cm
\begin{abstract}
The thermalisation rate for long wavelength fluctuations of the gauge
field in the abelian Higgs model is calculated from the imaginary part
of the self energy. The calculation is performed for both the symmetric
and symmetry broken phase of the theory.
\end{abstract}

\vskip 1cm

\renewcommand{\thefootnote}{\arabic{footnote}}
\setcounter{footnote}{0}

\section{Introduction}

The study of gauge theories at high temperatures has found
application in the description of a quark-gluon plasma in heavy ion
collisions and for the physics of the early universe. In recent years,
it has become apparent that conventional perturbation theory breaks
down when considering energy scales much smaller than the
temperature. In the case of hot QCD, this realisation led to the
development of a resummation scheme based on hard thermal
loops\cite{bratpis,bratpistherm}.

Recently in a paper by Elmfors, Enqvist and Vilja \cite{elm}, the
method of hard thermal loops was applied to the electroweak theory in
order to determine the thermalisation rate of fluctuations in the
Higgs field. In this paper, we use the same methods to study the
thermalisation rate of the gauge field in the abelian Higgs
model. This is a first step towards the calculation of the
thermalisation rate for a more realistic theory. The thermalisation
rate may be used to determine the time at which the gauge bosons
associated with a grand unified theory will `freeze out' in an
expanding universe. This may be of use in the subject of baryogenesis
in the early universe.

In the present paper we calculate the thermalisation rate $\gamma$
from the imaginary part of the temperature dependent Feynman self
energy $\Sigma$ via \cite{LvW,kapusta}
\begin{equation}
\gamma = -\frac{\mbox{Im}\Sigma(\omega,{\bf k})}{2\omega}
\end{equation}
where $\omega$ is the energy of a given mode.  We shall be calculating
the thermalisation rate of large scale fluctuations (that is at zero
3-momentum).  We use the real time formalism of thermal field
theory\cite{LvW,kapusta,NS,raybook} since it is straightforward to
extract the imaginary part of the self energy in this formalism
\cite{fujimoto}.

The paper is organised as follows. In section 2 we give a brief
overview of the abelian Higgs model and present the corrections to the
propagators and vertices in the hard thermal loop approximation. In
section 3 we use these results to calculate the thermalisation rate
both above and below the critical temperature at which the phase
transition occurs. Finally, we discuss the results in section 4.

\section{The Abelian Higgs model}
The abelian Higgs model consists of a complex scalar field minimally
coupled to a $U(1)$ gauge field. It is described by the Lagrangian
\begin{equation}
{\cal L} = -{\textstyle \frac{1}{4}} F^{\mu\nu}F_{\mu\nu} +
(D^{\mu}\Phi)^{*}(D_{\mu}\Phi) - \rho^{2}\Phi^{*}\Phi
-{\textstyle \frac{\lambda}{6}}(\Phi^{*}\Phi)^{2} \label{lag}
\end{equation}
with covariant derivative $D^{\mu}=\partial^{\mu}-\imath e A^{\mu}$
and abelian field strength tensor $F^{\mu\nu}=\partial^{\mu}
A^{\nu}-\partial^{\nu} A^{\mu}$. We choose $\rho^{2}<0$ so that
spontaneous symmetry breaking occurs. Expanding $\Phi$ about its
expectation value as
$$
\Phi(x)=\frac{1}{\surd{2}} \bigg{(} v+\phi(x)+\imath \chi(x) \bigg{)}
$$
we may rewrite Eqn.\ref{lag} as ${\cal L} = {\cal L}_{0} - V(A^{\mu},
\phi, \chi)$ where
\begin{eqnarray}
{\cal L}_{0} &= & -{\textstyle\frac{1}{4}} F^{\mu\nu} F_{\mu\nu} +
{\textstyle\frac{1}{2}} M^{2} A^{\mu} A_{\mu} \nonumber \\
& & + {\textstyle\frac{1}{2}}\partial^{\mu} \phi\partial_{\mu}\phi -
{\textstyle\frac{1}{2}} m^{2}\phi^{2} + {\textstyle\frac{1}{2}}
\partial_{\mu}\chi\partial^{\mu}\chi -
M \partial_{\mu} \chi A^{\mu}
\label{free}
\end{eqnarray}
and $V(A^{\mu},\phi ,\chi )$ is a potential term containing only cubic
and quartic terms. $M= ev$ and $m^{2}=\frac{1}{3}\lambda v^{2}$. We
assume $\lambda\sim e^{2}$.

The gauge is chosen so as to simplify the calculation. That is we
choose the gauge so as to both simplify the explicit form of the
propagators used and also to reduce the number of diagrams
contributing to hard thermal loops. To this end we choose to use the
t'Hooft gauge (the renormalisable $R_{\zeta}$ gauge with
$\zeta=1$).

We do not use the covariant gauge as the propagators in this gauge
must be resummed due to mixing between the photon and the unphysical
goldstone mode. As shown in \cite{epmgb}, this resummation leads to
real time propagators that contain derivatives of Dirac delta
functions. Although there is in principle no problem with using such
propagators, it is much simpler to use the renormalisable gauges which
do not need this kind of resummation.

The t'Hooft gauge is then chosen as it enables us to reduce the number
of diagrams contributing hard thermal loops. This is due to the
particularly simple form of the photon propagator.  The only
disadvantage of this choice of gauge is that Fadeev-Popov ghosts must
be included.

As a result of this choice of gauge, the following terms must be added
to the lagrangian.
\begin{equation}
{\cal L}_{g.f.} = -\frac{1}{2} \bigg{(} \partial_{\mu} A^{\mu} + M
\chi \bigg{)}^{2} - {\bar c} \bigg{(} \Box + M^{2} + eM\phi \bigg{)}
c
\end{equation}
where $c$ is the ghost field. As a result of this choice of gauge, the
Goldstone mode acquires a mass $M$ equal to that of the gauge boson.

\subsection{Leading order results}
The leading order corrections in the hard thermal loop approximation
are known in the case of the abelian higgs
model\cite{smilga,ueda,vokos,rebhan,rebsintra,frnktayl}. We shall not
repeat these calculations but merely quote the results. At high
temperatures, the vacuum expectation value(VEV) decreases from it's
zero temperature value, reaching zero at the critical temperature
$T_{C}$. The temperature dependant VEV $v_{\beta}$ is given by
\begin{equation}
v_{\beta}^{2}=v^{2}-\frac{T^{2}}{2}\bigg{(}
\frac{2}{3}+3\frac{e^{2}}{\lambda} \bigg{)}
\label{vev}
\end{equation}
The masses of the scalar and ghost sectors of the theory, are affected
only by the change in the VEV. That is the temperature dependant
masses $m_{\beta}$ and $M_{\beta}$ are defined by
\begin{eqnarray}
m_{\beta}^{2} &= &\frac{1}{3}\lambda v_{\beta}^{2} \nonumber \\
M_{\beta} &= & ev_{\beta}
\end{eqnarray}
However for the gauge boson there is a high temperature correction in
addition to that due to the temperature dependant VEV. This additional
contribution depends on the polarisation. That is there are three mass
corrections corresponding to the spatially transverse and longitudinal
polarisations and to the 4-vector longitudinal polarisation. These
masses are given by \cite{smilga}
\begin{equation}
\begin{array}{crcl}
\mbox{Transverse:} & \Pi_{T}(k_{0},{\bf k}) &= &M_{\beta}^{2} +
{\displaystyle\frac{e^{2}T^{2}}{6}}\bigg{\{} \frac{k_{0}^{2}}{{\bf k}^{2}} +
\bigg{(} 1 -\frac{k_{0}^{2}}{{\bf k}^{2}} \bigg{)}
\frac{k_{0}}{2{\bf k}} \ln\bigg{(} \frac{k_{0}+|{\bf k}|}{k_{0}-|{\bf
k}|} \bigg{)} \bigg{\}} \\
\nonumber\\
\mbox{Longitudinal:} & \Pi_{L}(k_{0},{\bf k}) &= &M_{\beta}^{2} +
{\displaystyle\frac{e^{2}T^{2}}{3}} \bigg{(}
1 - \frac{k_{0}^{2}}{{\bf k}^{2}} \bigg{)}
\bigg{\{} 1- \frac{k_{0}}{2|{\bf k}|} \ln\bigg{(} \frac{k_{0}+|{\bf
k}|} {k_{0}-|{\bf k}|}
\bigg{)} \bigg{\}} \label{asymmass}\\
\nonumber\\
\mbox{4-longitudinal:} & \Pi_{4}^{2} &= &M^{2}_{\beta}
\end{array}
\end{equation}
There are no hard thermal loop corrections to the vertices
\cite{rebhan,frnktayl}.

We shall be calculating the thermalisation rate at zero momentum. As
such we define $M_{T}$ to be the value of the self energy of the
transverse and longitudinal modes in this limit.
\begin{equation}
M_{T}^{2}=\Pi_{T}(k_{0},{\bf 0}) = \Pi_{L}(k_{0},{\bf 0}) =
M_{\beta}^{2} + \frac{e^{2}T^{2}}{9}
\label{zmass}
\end{equation}

Above the critical temperature, the symmetry is restored. There is now
no distinction between the higgs and goldstone fields. The masses for
the particles are now
\begin{eqnarray}
\mbox{Higgs/Goldstone:} & m^{2}=\frac{1}{12}(T^{2}-T_{C}^{2})
\{ \frac{2}{3}\lambda + 3e^{2} \} &\nonumber\\
\nonumber\\
\mbox{Photon:} &\mbox{As in Eqns.\ref{asymmass}-\ref{zmass} but with }
M_{\beta}=0
\label{symmass}
\end{eqnarray}

\section{Calculation of the thermalisation rate}
To calculate the thermalisation rate we need the imaginary part of the
self-energy\cite{weldon}. In the real time formalism, the imaginary
part of the Feynman self energy is most easily calculated from the
$1$-$2$ component of the self energy via the relation
\cite{fujimoto,LvW}
\begin{equation}
\mbox{Im} \Sigma=-\frac{e^{-\frac{\beta |k_{0}|}{2}}}{2n(|k_{0}|)}
\mbox{Im} \Sigma_{12}
\end{equation}
where $n(X)$ is the Bose-Einstein distribution. We are working in the
symmetric version of the real time formalism (that is
$\sigma=\frac{1}{2}$ in the notation of Landsman and van Weert
\cite{LvW}).

\subsection{Resummed propagators}
To proceed with the calculation we need the explicit form for the
$1$-$2$ component of the resummed propagators. For the scalar and
ghost sector of the theory, this is easy due to the simple form of the
hard thermal loop mass correction. However care must be taken with the
photon propagator due to the complicated momentum dependance of the
self energies. The $1$-$2$ propagators for the higgs and goldstone
modes are given by
\begin{eqnarray}
\mbox{Higgs:}&\imath\Delta^{H}_{12}(k)&=2\pi e^{\frac{\beta
|k_{0}|}{2}}n(|k_{0}|) \delta (k^{2}-m^{2}_{\beta}) \\
\nonumber\\
\mbox{Goldstone:} &\imath\Delta^{G}_{12}(k)&=2\pi e^{\frac{\beta
|k_{0}|}{2}}n(|k_{0}|) \delta (k^{2}-M^{2}_{\beta})
\end{eqnarray}
We must take care with the photon propagator since for $k_{0}^2 <{\bf
k}^{2}$ the self energies of the transverse and longitudinal modes
acquire an imaginary contribution. By dividing the propagator up
into it's three polarisations we can treat each case separately.
\begin{equation}
\imath\Delta^{\mu\nu}_{12}(k)= P^{\mu\nu}_{T} \imath\Delta^{T}_{12}(k)
+P^{\mu\nu}_{L}\imath\Delta^{L}_{12}
+ \frac{k^{\mu}k^{\nu}}{k^{2}} \imath\Delta^{4}_{12}(k)
\end{equation}
where $P_{T}^{\mu\nu}$ and $P_{L}^{\mu\nu}$ are the transverse and
longitudinal projection operators. In the rest frame of the heat bath
they are given by
\begin{eqnarray}
P_{T}^{00}(k) &= &P_{T}^{0i}(k)\hspace{6pt}=\hspace{6pt}P_{T}^{i0}(k)
\hspace{6pt}=\hspace{6pt}0 \nonumber\\
P_{T}^{ij}(k) &= &-\bigg{(} \delta^{ij} - \frac{k^{i} k^{j}}{{\bf
k}^{2}}\bigg{)} \nonumber\\
P_{L}^{\mu\nu}(k) &= &\bigg{(} g^{\mu\nu}-\frac{k^{\mu}k^{\nu}}{k^{2}}
\bigg{)} - P_{T}^{\mu\nu}(k)
\end{eqnarray}
{}From this we find that
$\Delta^{4}_{12}(k)=-\Delta^{G}_{12}(k)$. To calculate
$\Delta^{T}_{12}$ and $\Delta^{L}_{12}$ we use the relation
\begin{equation}
\Delta_{12}=e^{\frac{\beta |k_{0}|}{2}} n(|k_{0}|) \bigg{(} \Delta_{F}
-\Delta^{*}_{F} \bigg{)}
\end{equation}
where in $\Delta_{F}$ we replace $k_{0}$ by $k_{0}+\imath\varepsilon$
($k_{0}-\imath\varepsilon$) when $k_{0}$ is greater than (less than)
zero. Using this we find that the transverse propagator is given by
\begin{equation}
\begin{array}{llll}
\imath\Delta^{T}_{12}(k_{0},{\bf k})&=
&-2\pi e^{\frac{\beta |k_{0}|}{2}} n(|k_{0}|)
\delta(k^{2}-\Pi_{T}(k_{0},{\bf k})) &k_{0}^{2}>{\bf k}^{2}
\\
\\ &= & -2\pi e^{\frac{\beta |k_{0}|}{2}} n(|k_{0}|)
{\displaystyle
\frac{2M_{TC}^{2}(k_{0},{\bf k})} {\bigg{(}k^{2}- \mbox{Re}(\Pi_{T}(k_{0},{\bf
k}))\bigg{)}^{2}+M_{TC}^{4}(k_{0}, {\bf k})}} & k_{0}^{2}<{\bf k}^{2}
\end{array} \label{tprop}
\end{equation}
where $\Pi_{T}$ is given by Eqn.\ref{asymmass} and $M_{TC}$ is
\begin{equation}
M^{2}_{TC}(k_{0},{\bf k}) = \pi \frac{e^{2}T^{2}}{12}
\frac{|k_{0}|}{|{\bf k}|} \bigg{(} 1-\frac{k_{0}^{2}}{{\bf k}^{2}} \bigg{)}
\end{equation}
Similarly, the longitudinal propagator is given by
\begin{equation}
\begin{array}{llll}
\imath\Delta^{L}_{12}(k_{0},{\bf k})&=
&-2\pi e^{\frac{\beta |k_{0}|}{2}} n(|k_{0}|)
\delta(k^{2}-\Pi_{L}(k_{0},{\bf k})) &k_{0}^{2}>{\bf k}^{2} \\
\\
&= & -2\pi e^{\frac{\beta |k_{0}|}{2}} n(|k_{0}|)
{\displaystyle
\frac{2 M^{2}_{LC}(k_{0},{\bf k})}{\bigg{(} k^{2} - \mbox{Re}(\Pi_{L}(k_{0},
{\bf k})) \bigg{)}^{2} + M_{LC}^{4}(k_{0},{\bf k})}}
& k_{0}^{2}<{\bf k}^{2}
\end{array}
\end{equation}
with $\Pi_{L}$ given by Eqn.\ref{asymmass} and $M_{LC}$ is
\begin{equation}
M^{2}_{LC}(k_{0},{\bf k}) = -\pi \frac{e^{2}T^{2}}{6}
\frac{|k_{0}|}{|{\bf k}|} \bigg{(} 1-\frac{k_{0}^{2}}{{\bf k}^{2}} \bigg{)}
\label{lprop}
\end{equation}

\subsection{The calculation}
\begin{figure}
\setlength{\unitlength}{2pt}
\begin{center}
$
-\imath\Sigma^{\mu\nu}_{12}(p)=
\begin{picture}(60,30)(-20,-3)
\multiput(-14,0)(8,0){2}{\oval(4,4)[t]}
\multiput(-10,0)(8,0){2}{\oval(4,4)[b]}
\multiput(24,0)(8,0){2}{\oval(4,4)[t]}
\multiput(28,0)(8,0){2}{\oval(4,4)[b]}
\put(11,11){\circle*{3}}
\put(11,-11){\circle*{3}}
\put(11,0){\oval(22,22)[t]}
\put(2,-4){\oval(4,4)[tr]}
\multiput(4,-4)(-0.1,-0.2){5}{\line(1,0){0.1}}
\multiput(3.5,-5)(-0.1,-0.1){5}{\line(1,0){0.1}}
\multiput(3,-5.5)(-0.1,-0.2){5}{\line(1,0){0.1}}
\put(4.5,-6.5){\oval(4,4)[bl]}
\multiput(4.5,-8.5)(0.2,0.1){5}{\line(1,0){0.1}}
\multiput(5.5,-8)(0.1,0.1){5}{\line(1,0){0.1}}
\multiput(6,-7.5)(0.2,0.1){5}{\line(1,0){0.1}}
\put(7,-9){\oval(4,4)[tr]}
\put(11,-9){\oval(4,4)[b]}
\put(15,-9){\oval(4,4)[tl]}
\multiput(15,-7)(0.2,-0.1){5}{\line(1,0){0.1}}
\multiput(16,-7.5)(0.1,-0.1){5}{\line(1,0){0.1}}
\multiput(16.5,-8)(0.2,-0.1){5}{\line(1,0){0.1}}
\put(17.5,-6.5){\oval(4,4)[br]}
\multiput(19.5,-6.5)(-0.1,0.2){5}{\line(1,0){0.1}}
\multiput(19,-5.5)(-0.1,0.1){5}{\line(1,0){0.1}}
\multiput(18.5,-5)(-0.1,0.2){5}{\line(1,0){0.1}}
\put(20,-4){\oval(4,4)[tl]}
\put(20,0){\oval(4,4)[br]}
\end{picture}
+
\begin{picture}(50,30)(-20,-3)
\multiput(-14,0)(8,0){2}{\oval(4,4)[t]}
\multiput(-10,0)(8,0){2}{\oval(4,4)[b]}
\multiput(24,0)(8,0){2}{\oval(4,4)[t]}
\multiput(28,0)(8,0){2}{\oval(4,4)[b]}
\put(11,11){\circle*{3}}
\put(11,-11){\circle*{3}}
\put(11,0){\oval(22,22)[t]}
\put(0,0){\line(0,-1){1}}
\multiput(0,-1)(0.04,-0.2){5}{\line(1,0){0.1}}
\multiput(2.5,-7.1)(-0.1,0.15){6}{\line(1,0){0.1}}
\multiput(2.5,-7.1)(0.1,-0.1){14}{\line(1,0){0.1}}
\multiput(3.9,-8.5)(0.15,-0.1){6}{\line(1,0){0.1}}
\multiput(10,-11)(-0.2,0.04){5}{\line(1,0){0.1}}
\put(10,-11){\line(1,0){2}}
\multiput(12,-11)(0.2,0.04){5}{\line(1,0){0.1}}
\multiput(19.5,-7.1)(0.1,0.15){6}{\line(1,0){0.1}}
\multiput(19.5,-7.1)(-0.1,-0.1){14}{\line(1,0){0.1}}
\multiput(18.1,-8.5)(-0.15,-0.1){6}{\line(1,0){0.1}}
\multiput(22,-1)(-0.04,-0.2){5}{\line(1,0){0.1}}
\put(22,0){\line(0,-1){1}}
\end{picture}
$
\end{center}
\caption{Feynman diagrams contributing to Im$\Sigma_{12}^{\mu\nu}$
(Dot denotes hard thermal loop corrected propagator)}

\label{feyndiag}
\end{figure}
There are only two diagrams contributing to $\Sigma_{12}$. They are
shown in Fig.\ref{feyndiag}. From these diagrams, we find that
$\Sigma_{12}^{\mu\nu}$ is given by
\begin{equation}
\Sigma_{12}^{\mu\nu}(p_{0},{\bf p})=-\imath e^{2} \int d^{4}k
\bigg{\{} 4M_{\beta}^{2}\imath\Delta_{12}^{\mu\nu}(k)
+ (p-2k)^{\mu}(p-2k)^{\nu}
\imath\Delta^{G}_{12}(k) \bigg{\}} \imath\Delta^{H}_{12}(p-k)
\label{imsig}
\end{equation}
Using this we may now calculate the imaginary part of the transverse
and longitudinal self-energies.
\begin{eqnarray}
\mbox{Im}\Sigma_{12}^{T}(p_{0},{\bf p}) &= & \frac{1}{2}
P^{T}_{\mu\nu}(p) \mbox{Im} \Sigma^{\mu\nu}_{12}(p_{0},{\bf p})
\nonumber\\
&= & -2e^{2}M_{\beta}^{2} P^{T}_{\mu\nu}(p)
\int \frac{d^{4}k}{(2\pi)^{4}} \imath\Delta^{H}_{12}(p-k)
\bigg{\{} P_{T}^{\mu\nu}(k) \imath\Delta^{T}_{12}(k) +
P_{L}^{\mu\nu}(k) \imath\Delta^{L}_{12}(k) \bigg{\}} \nonumber\\
&& \hspace{.75in} -\frac{e^{2}}{2} \int \frac{d^{4}k}{(2\pi)^{4}}
\imath\Delta_{12}^{H}(p-k)\imath\Delta_{12}^{G}(k). \nonumber\\
&& \hspace{1in} . P^{T}_{\mu\nu}(p) \bigg{\{} (p-2k)^{\mu}
(p-2k)^{\nu} -  4M^{2}_{\beta} \frac{k^{\mu} k^{\nu}}{k^{2}}
\bigg{\}}
\label{trans}
\end{eqnarray}
We have divided up the integral in Eqn.\ref{trans} into two
sections. The first term contains the contribution from the physical
modes of the photon propagator. The second term contains the
contribution from the unphysical modes of the photon propagator and
the contribution from the Goldstone propagator. Evaluating the second
term in Eqn.\ref{trans} (which we shall label as $\Sigma^{U}_{12}$), we
find
\begin{equation}
\Sigma^{U}_{12}= 2e^{2} \int \frac{d^{4}k}{(2\pi)^{4}}
\imath\Delta^{H}_{12}(p-k) \imath\Delta^{G}_{12}(k)
\frac{(k^{2}-M^{2}_{\beta})}{k^{2}}
\bigg{(} {\bf k^{2}-\frac{(p.k)^{2}}{p^{2}}} \bigg{)} = 0
\end{equation}
due to the Dirac delta function in $\Delta^{G}_{12}(k)$.  We therefore
find that the only contribution to $\mbox{Im}\Sigma_{T}$ comes from
the physical modes of the theory. This suggests that our calculation is
gauge invariant.
\begin{equation}
\mbox{Im}\Sigma_{T}=e^{2}M_{\beta}^{2}
\frac{e^{\frac{\beta |p_{0}|}{2}}}{n(|p_{0}|)}
\int \frac{d^{4}k}{(2\pi)^{4}} \imath\Delta^{H}_{12}(p-k)
P^{\mu\nu}_{T}(p) \bigg{\{}
P_{\mu\nu}^{T}(k)\imath\Delta_{12}^{T}(k) +
P_{\mu\nu}^{L}(k)\imath\Delta_{12}^{L}(k) \bigg{\}}
\label{imsigm}
\end{equation}
Similarly, when calculating $\mbox{Im}\Sigma_{L}$, the unphysical
contributions cancel out.

We now evaluate Eqn.\ref{imsigm} at $p_{0}=M_{T},{\bf p=0}$. The
integration splits up into two regions corresponding to the two
regions over which $\Delta^{T}_{12}$ and $\Delta^{L}_{12}$ are
defined (see Eqns.\ref{tprop}-\ref{lprop}). That is we split up the
integral into the regions $k^{2}>0$ and $k^{2}<0$. There is no
contribution from the region $k^{2}>0$ since at zero momentum we
cannot simultaneously satisfy both Dirac delta functions in
Eqn.\ref{imsigm}.

The region $k^{2}<0$ does contribute. Performing all but the final
momentum integration we find
\begin{eqnarray}
\mbox{Im}\Sigma_{T}&= &-\frac{e^{2}M_{\beta}^{2}}{3\pi^{2}}
\int_{k>|M_{T}-\omega|} dk \frac{k^{2}}{\omega}
\frac{n(|M_{T}-\omega|)n(\omega)}{n(M_{T})}
e^{\frac{\beta}{2}(|M_{T}-\omega|+\omega-M_{T})}. \label{Yuk}\\
&& \bigg{[} \frac{2 M_{TC}(M_{T}-\omega,k)}{\bigg{(}
(M_{T}-\omega)^{2}-k^{2}
- \Pi_{T}(M_{T}-\omega,k) \bigg{)}^{2} +
M_{TC}^{4}(M_{T}-\omega,k)} \nonumber\\
&&
+ \frac{(M_{T}-\omega)^{2}}{(M_{T}-\omega)^{2}-k^{2}}
\frac{M_{LC}(M_{T}-\omega,k)}
{\bigg{(} (M_{T}-\omega)^{2}-k^{2} - \Pi_{L}(M_{T}-\omega,k)
\bigg{)}^{2}
+ M_{LC}^{4}(M_{T}-\omega,k)} \bigg{]} \nonumber
\end{eqnarray}
Similarly $\mbox{Im}\Sigma_{L}$ may be evaluated. At zero momentum we
find $\mbox{Im}\Sigma_{T}=\mbox{Im}\Sigma_{L}$. This is to be expected
since at zero momentum one cannot distinguish between transverse and
longitudinal polarisations.

We must now consider the high temperature limit. Examining the
temperature dependant terms from Eqn.\ref{Yuk}, and using the Mellin
transformation\cite{LvW,GRET} to extract the high temperature behaviour we find
\begin{eqnarray}
\frac{n(|M_{T}-\omega|)n(\omega)}{n(M_{T})}
e^{\frac{\beta}{2}(|M_{T}-\omega|+\omega-M_{T})} &= &
\left\{
\begin{array}{cc}
1 + n(\omega) + n(M_{T}-\omega) & M_{T}>\omega \\
n(\omega-M_{T})-n(\omega) & M_{T}<\omega
\end{array}
\right. \\
&\sim & \frac{M_{T} T}{\omega |M_{T}-\omega|}
\end{eqnarray}
We therefore find that to $O(e^{2}T)$, the thermalisation rate is
given by
\begin{equation}
\gamma({\bf p}=0) =
\frac{e^{2}T}{24\pi}.
A({\textstyle\frac{T}{T_{C}},\frac{\lambda}{e^{2}}})
\label{ans}
\end{equation}
where $A(\frac{T}{T_{C}},\frac{\lambda}{e^{2}})$ is a dimensionless
function of the ratio $\frac{\lambda}{e^{2}}$ and the ratio of the
temperature to the critical temperature. We have chosen the factor of
$24\pi$ in Eqn.\ref{ans} so as to compare the result with the
thermalisation rate above the critical temperature (See section 3.3).

\begin{eqnarray}
A(\frac{T}{T_{C}},\frac{\lambda}{e^{2}}) &= &\int_{{\bar k}>|{\bar
M_{T}} - {\bar \omega}|} d{\bar k} \frac{2{\bar M_{\beta}^{2}}}
{3{\bar \omega^{2}}{\bar k}}.
\label{ans2}
\\
&& \hspace{.2in}
.\bigg{[}
\frac{{\bar k}^{2}-({\bar M_{T}}-{\bar \omega})^{2}}
{\bigg{(} ({\bar M_{T}}-{\bar \omega})^{2}-{\bar k}^{2} - {\bar
\Pi_{T}}({\bar M_{T}} - {\bar \omega},{\bar k}) \bigg{)}^{2} +
{\bar M_{TC}}^{4}({\bar M_{T}}-{\bar \omega},{\bar k})} \nonumber \\
&& \hspace{.21in}
+ \frac{({\bar M_{T}}-{\bar \omega})^{2}}
{\bigg{(} ({\bar M_{T}}-{\bar \omega})^{2}-{\bar k}^{2} - {\bar
\Pi_{L}}({\bar M_{T}} - {\bar \omega},{\bar k}) \bigg{)}^{2} +
{\bar M_{LC}}^{4}({\bar M_{T}}-{\bar \omega},{\bar k})} \bigg{]}
\nonumber
\end{eqnarray}
where the overline denotes that the variable has been scaled by a
factor of $eT$. That is ${\bar k}=\frac{k}{eT}$. Note that
$A(\frac{T}{T_{C}},\frac{\lambda}{e^{2}})$ is positive definite.

\begin{figure}
{\centerline{\psfig{file=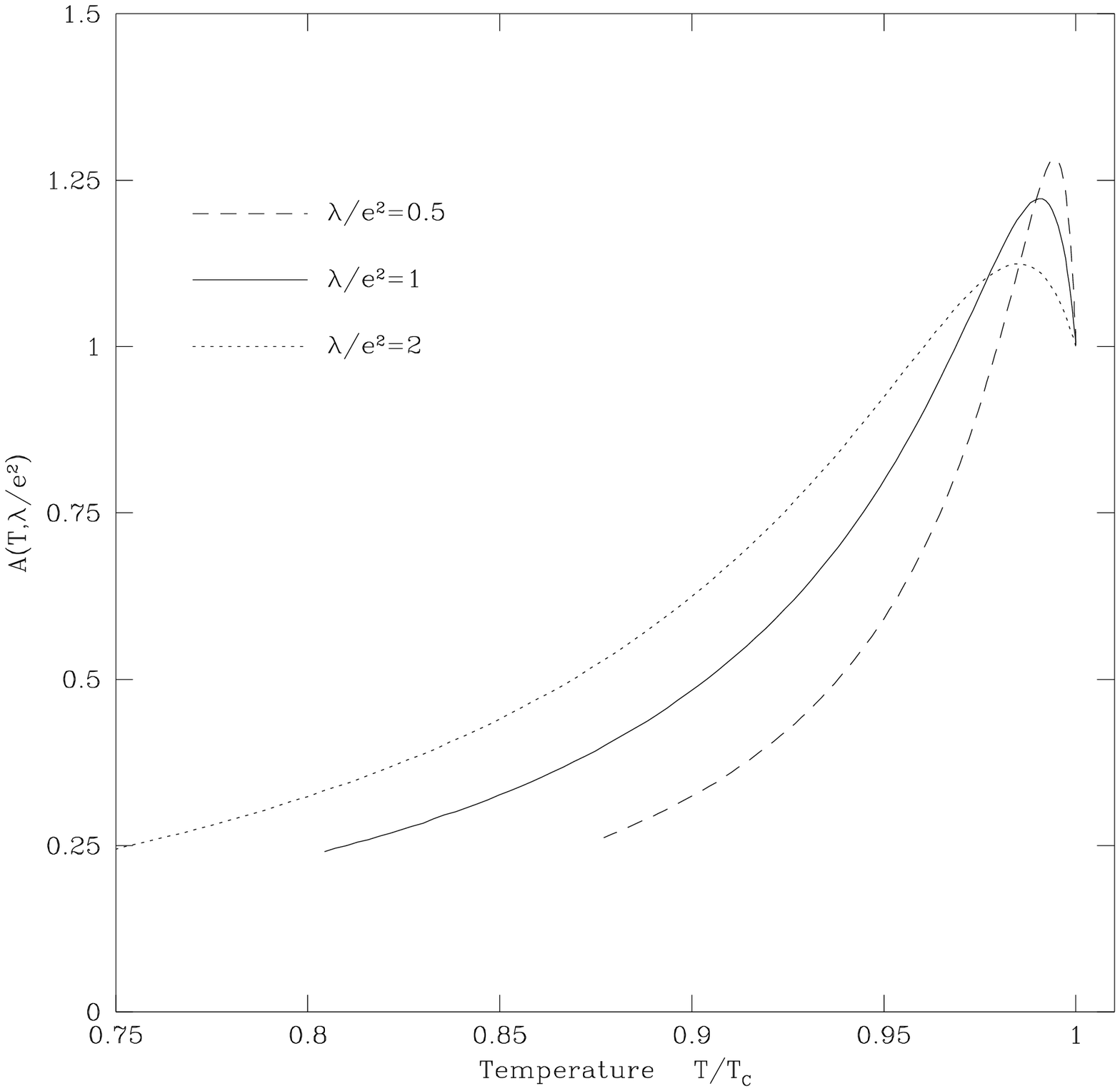,width=6in}}}
\caption{Temperature dependance of the constant
$A(T,\frac{\lambda}{e^{2}})$}
\label{aplo}
\end{figure}

This expression depends on the validity of the hard thermal loop
approximation. We are assuming that all the masses are smaller than
the temperature by at least one factor of the coupling constant $e$.
That is we should only trust the results for $\frac{v_{\beta}}{T}\leq
1$. We should also be careful that the constant
$A(\frac{T}{T_{C}},\frac{\lambda}{e^{2}}) \geq e$.  In addition to
this we must also be sure that we do not vary the value of
$\frac{\lambda}{e^{2}}$ so as to invalidate our assumption that
$\lambda\sim e^{2}$.  The temperature dependance of
$A(\frac{T}{T_{C}},\frac{\lambda}{e^{2}})$ is shown in Fig.\ref{aplo}
for a range of values of $\frac{\lambda}{e^{2}}$. It can be seen that
near the critical temperature
$A(\frac{T}{T_{C}},\frac{\lambda}{e^{2}})\sim 1$. However as the
temperature decreases, $A(\frac{T}{T_{C}},\frac{\lambda}{e^{2}})$
becomes smaller. This is due to the increasing size of $M_{\beta}$
reducing the effect of Landau damping.

\begin{figure}
{\centerline{\psfig{file=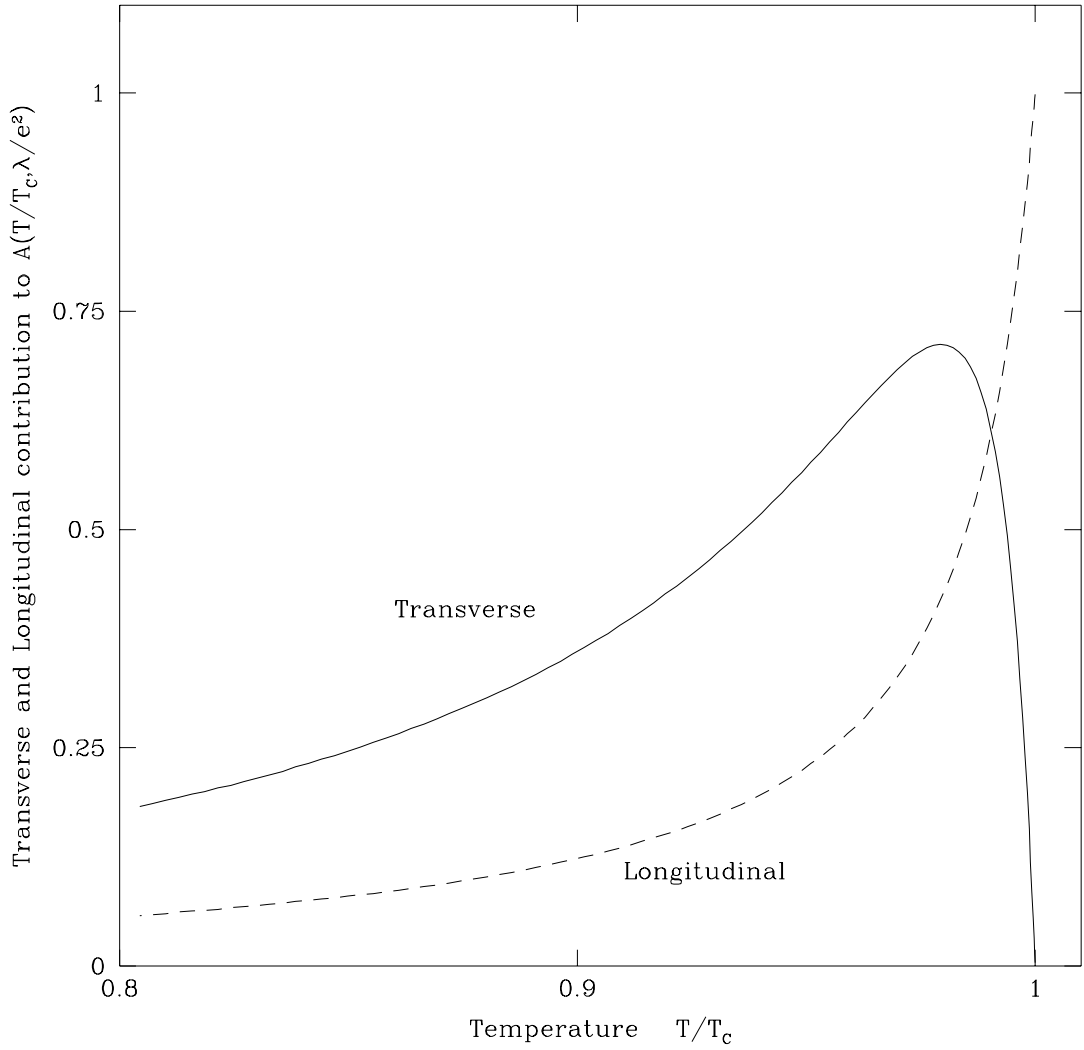,width=6in}}}
\caption{Contribution to $A(\frac{T}{T_{C}},\frac{\lambda}{e^{2}})$
from the transverse and longitudinal parts of Eqn.28}
\label{dectlplo}
\end{figure}

One might expect that since $A(\frac{T}{T_{C}},\frac{\lambda}{e^{2}})
\propto M_{\beta}^{2}$, we should expect that
$A(\frac{T}{T_{C}},\frac{\lambda}{e^{2}})$ vanishes at the critical
temperature. However, this is not the case. In Fig.\ref{dectlplo} we
show the contribution to $A(\frac{T}{T_{C}},\frac{\lambda}{e^{2}})$
from the two terms in Eqn.\ref{ans2}. These correspond to the
contributions from the transverse and longitudinal parts of the photon
propagator. It can be seen that near the critical temperature the
longitudinal contribution dominates. This is due to the behaviour of
$\Pi_{T}$ and $\Pi_{L}$ near the light cone. As we approach the critical
temperature, $\Pi_{L}$ becomes small around the light cone and can compensate
for the smallness of $M_{\beta}$ in the integrand of Eqn.\ref{ans2}.
However, $\Pi_{T}$ does not vanish and so the contribution to
$A(\frac{T}{T_{C}},\frac{\lambda}{e^{2}})$ from the first term in
Eqn.\ref{ans2} vanishes as we approach the critical temperature.

\subsection{Above the critical temperature}
\begin{figure}
{\centerline{\psfig{file=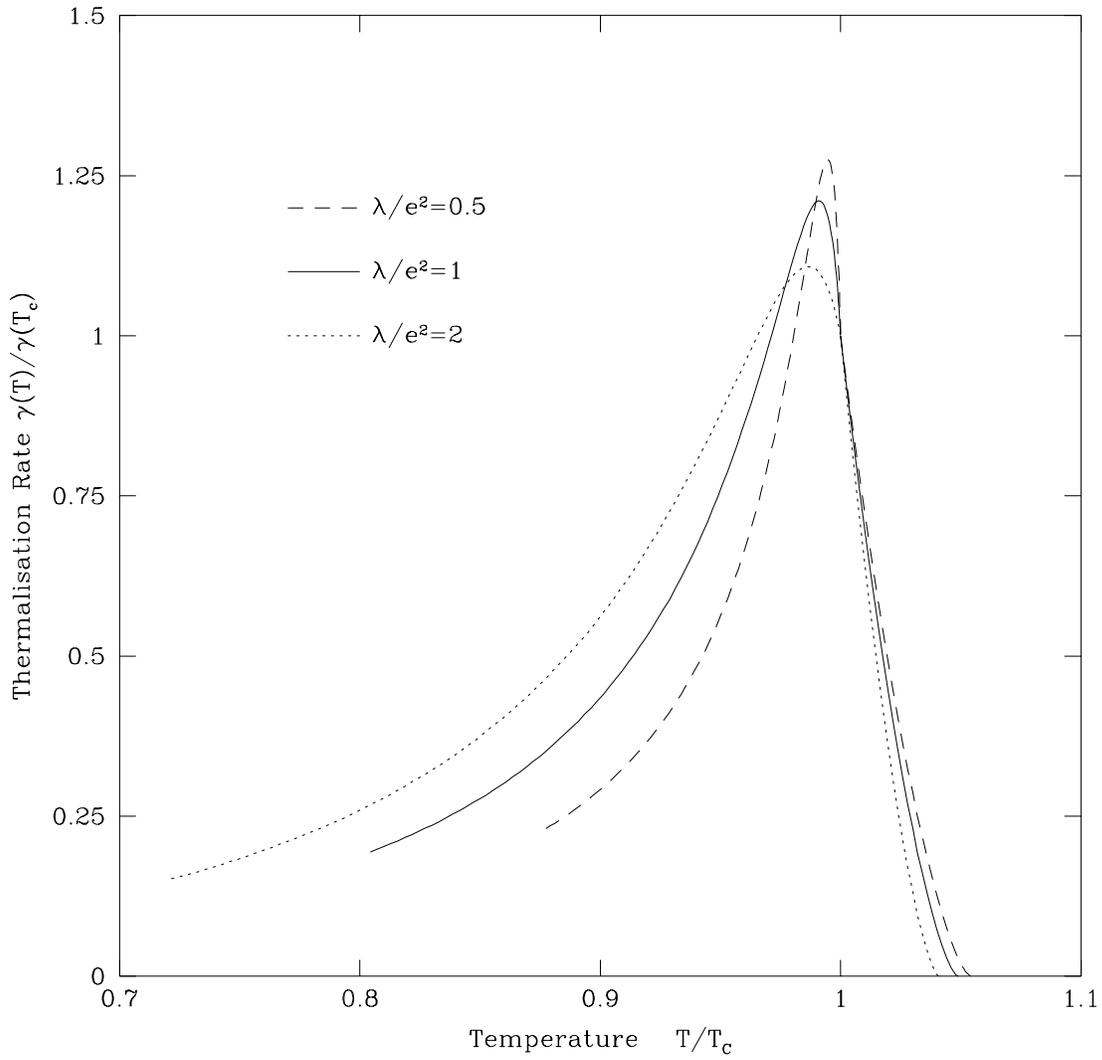,width=6in}}}
\caption{Plot of thermalisation rate versus temperature}
\label{decplo}
\end{figure}

In the symmetric phase, only the second Feynman diagram in
Fig.\ref{feyndiag} contributes to $\Sigma_{12}$. From this we find
that the imaginary part of the self energy at zero momentum is given
by
\begin{equation}
\mbox{Im} \Sigma_{T}(p_{0},{\bf p}) = -e^{2} \frac{e^
{\frac{\beta |p_{0}|}{2}}}{2n(|p_{0}|)}
\int \frac{d^{4}k}{(2\pi)^{4}} P_{T}^{\mu\nu}(p) \big{[}
p-2k \big{]}_{\mu} \big{[} p-2k \big{]}_{\nu}
\imath\Delta^{H}_{12}(p-k) \imath\Delta^{G}_{12}(k)
\end{equation}
Evaluating this we find that the thermalisation rate at zero momentum
above the critical temperature is given by
\begin{equation}
\gamma({\bf p}=0)=\frac{e^{2}T}{24\pi}\bigg{(}
1-\frac{4m^{2}}{M_{T}^{2}} \bigg{)}^{\frac{3}{2}}
\hspace{.3in} M_{T}>2m
\label{symdec}
\end{equation}
where $M_{T}=\frac{eT}{3}$ is the value of the transverse or
longitudinal self energy at zero momentum and m is defined by
Eqn.\ref{symmass}. For $M_{T}>2m$, this diagram does not contribute to
the imaginary part of the self energy and the thermalisation rate is
of a higher order in the perturbation expansion associated with hard
thermal loops. Solving the equation $M_{T}=2m$ we find that
Eqn.\ref{symdec} is valid in the temperature region
\begin{equation}
1 \leq \frac{T}{T_{C}} \leq
\sqrt{\frac{2\lambda+9e^{2}}{2\lambda+8e^{2}}}
\end{equation}
It should be noted that the value of the thermalisation rate at the
critical temperature is independent of whether we approach the
critical temperature from above or below. The decay rate as a function
of temperature is shown in Fig.\ref{decplo}. This result is consistent
with the thermalisation rate for scalar electrodynamics \cite{rebhan}.

\section{Conclusions}
We have managed to calculate the thermalisation rate at zero momentum
for the gauge boson in the abelian Higgs model. The next step is to
consider extending this work to a more realistic theory that contains
fermions and utilises a larger gauge symmetry. While studying such a
theory is more involved, there are in principle no additional problems
in calculating the thermalisation rate.

At the order to which we work a second order phase transition is
indicated. However near the phase transition $v_{\beta}$ is small and
higher order corrections become important. The quantitative results
may change but more importantly the qualitative picture may change and
the theory may have a first order phase transition. However we may
trust our results away from the critical temperature.


\end{document}